\documentclass[twocolumn,showpacs,floatfix,eqsecnum,prb]{revtex4}

\usepackage[dvips]{epsfig}
\usepackage{amssymb}

\usepackage{float}

\begin{document}

\title{Surface States of the Topological Insulator Bi$_{1-x}$ Sb$_x$}

\author{Jeffrey C.Y. Teo, Liang Fu and C.L. Kane}
\affiliation{Dept. of Physics and Astronomy, University of
Pennsylvania, Philadelphia, PA 19104}

\begin{abstract}
We study the electronic surface states of the semiconducting alloy
bismuth antimony (Bi$_{1-x}$Sb$_x$). Using a phenomenological tight
binding model we show that the Fermi surface for the 111 surface
states encloses an odd number of time reversal invariant momenta
(TRIM) in the surface Brillouin zone.  This confirms that the alloy is a
strong topological insulator in
the $(1;111)$ $\mathbb{Z}_2$ topological class. We go on to develop general arguments which
show that spatial symmetries lead to additional topological structure
of the bulk energy bands, and impose further constraints on the
surface band structure. Inversion symmetric band structures are
characterized by 8 $\mathbb{Z}_2$ ``parity invariants", which include
the 4 $\mathbb{Z}_2$ invariants defined by time reversal symmetry.
The extra invariants determine the ``surface fermion parity", which
specifies which surface TRIM are enclosed by an odd number of
electron or hole pockets.  We provide a simple proof of this result,
which provides a direct link between the surface state structure and
the parity eigenvalues characterizing the bulk.  Using this result we
make specific predictions for the surface state structure for several
faces of Bi$_{1-x}$Sb$_x$.  We next show that mirror invariant
band structures are characterized by an integer ``mirror Chern
number", $n_{\cal M}$, which further constrains the surface states.  We show
that the sign of $n_{\cal M}$ in the topological insulator phase of
Bi$_{1-x}$Sb$_x$ is related to a previously unexplored $\mathbb{Z}_2$ parameter
in the L point ${\bf k}\cdot{\bf p}$ theory of pure
bismuth, which we refer to as the ``mirror chirality", $\eta$.  The value of
$\eta$ predicted by the tight binding model for bismuth
disagrees with the value predicted by a more fundamental
pseudopotential calculation. This explains a subtle disagreement
between our tight binding surface state calculation and previous first
principles calculations of the surface states of bismuth.  This
suggests that the tight binding parameters in the Liu Allen model of
bismuth need to be reconsidered.  Implications for existing and
future angle resolve photoemission (ARPES) experiments and spin polarized
ARPES experiments will be discussed.

\end{abstract}

\maketitle

\section{Introduction}

A topological insulator is a material with a bulk electronic
excitation gap generated by the spin orbit interaction, which is
topologically distinct from an ordinary insulator
\cite{km1,km2,bernevig,roy1,moore,roy2,fkm,bhz}.
This distinction,
characterized by a $\mathbb{Z}_2$ topological invariant, necessitates the
existence of gapless electronic states on the sample boundary. In two
dimensions, the topological insulator is a quantum spin Hall
insulator\cite{km1,km2,bernevig,bhz},
which is a close cousin of the integer quantum Hall state.
The edge states predicted for this phase have recently been observed
in transport experiments on HgCdTe quantum wells\cite{molenkamp}.  In three
dimensions there are four $\mathbb{Z}_2$ invariants characterizing a
time reversal invariant band
structure\cite{moore,roy2,fkm}.  One of these distinguishes a strong topological
insulator, which is robust in the presence of disorder.  The strong topological insulator
is predicted to have surface states whose Fermi
surface encloses an odd number of Dirac points and is associated with
a Berry's phase of $\pi$.  This defines a topological metal surface
phase, which is predicted to have novel electronic properties\cite{fkm,fukane3,qizhang}

In Ref. \onlinecite{fukane2} we predicted that the
semiconducting alloy Bi$_{1-x}$ Sb$_x$ is a strong topological
insulator using a general argument based on the inversion symmetry of
bulk crystalline Bi and Sb.  The surface states of Bi have been studied for several years.
Experimentally there are several photoemission studies of Bi crystals
and films which have probed the surface states
\cite{review,patthey,hofmann2,ast1,ast2,pascual,koroteev1,hofmann1,hirahara1,hirahara2}.   There are fewer
studies of Bi$_{1-x}$ Sb$_x$\cite{hochst}, but in very recent work, Hsieh et al. \cite{hasan}
have mapped the (111) surface states, and verified the
topological structure predicted for a strong topological insulator.

First principles calculations provide a clear picture of the surface state
structure of Bi\cite{koroteev1,hofmann1, hirahara1,hirahara2,koroteev2},
which captures many of the experimental features,
including their spin structure\cite{hirahara2}.  For the alloy, Bi$_{1-x}$ Sb$_x$,
one expects the surface states to evolve smoothly from Bi, at least
for small $x$.  The alloy presents two difficulties for these
calculations, though.    First, since these calculations must be done
on relatively thin slabs, features near the small band gap are
inaccessible because finite size quantization mixes the bulk and
surface states.  Moreover, describing the alloy would require some
kind of mean field treatment of the substitutional disorder.

In this paper we study the surface states of Bi$_{1-x}$ Sb$_x$ first
by developing a phenomenological tight binding model which can be solved
numerically and then by developing general arguments that exploit spatial
symmetries and explain a
number of model independent features of the surface states.
Our phenomenological tight binding model is based on an interpolation of a
model developed by Liu and Allen\cite{liuallen}.  This model has the advantage that
it can be solved in a semi infinite geometry, which allows the
surface state features near the small band gap to be calculated. Our
aim is not to perform a quantitatively accurate calculation of the
surface states, but rather to provide a concrete calculation in which
robust, model independent features of the surface states can be
identified and characterized.   Here we list our main conclusions:

(1)   We find that the Fermi surface of the 111 surface of Bi$_{1-x}$ Sb$_x$
consists of an electron pocket centered around the $\bar\Gamma$
point and six elliptical hole pockets centered a point in between
$\bar\Gamma$ and the $\bar M$ point.  (Here the bar refers to symmetry
points in the 111 surface Brillouin zone).  This is similar to the surface states in
Bi.  Unlike the alloy, however, Bi has bulk states at the Fermi energy:
hole states near $\bar\Gamma$ and electron states near $\bar M$.

This calculation verifies the topological structure of the surface states
predicted in Ref. \onlinecite{fukane2}.  In that work we showed that the
four $\mathbb{Z}_2$ invariants $(\nu_0;\nu_1\nu_2\nu_3)$ characterizing
the valence bands of pure Bi and Sb are $(0;000)$ and $(1;111)$
respectively.  The semiconducting alloy Bi$_{1-x}$ Sb$_x$ was argued
to be in the same class as Sb, which is a strong topological
insulator.  These invariants determine the number of surface bands
crossing the Fermi energy modulo 2 between each pair of time
reversal invariant momenta (TRIM) in the surface Brillouin zone.
Specifically, it predicts that for the 111 surface an odd number of Fermi surface
lines separate the $\bar\Gamma$ point from the three equivalent $\bar M$ points.
This is consistent with both our calculation and with experiment\cite{hasan}.

(2)  We will show that for crystals with inversion symmetry
there is additional topological structure in the bulk band structure,
which further constrains the surface band structure.  At each of the
8 TRIM, $\Gamma_i$ in the bulk Brillioun zone, the product of the parity
eigenvalues of the occupied bands defines a {\it parity invariant}
$\delta(\Gamma_i)$, which is a topological invariant in the space of
inversion symmetric Hamiltonians.  The four $\mathbb{Z}_2$ invariants, which
require only time reversal symmetry are determined by these 8 signs,
and determine the number of Fermi surface lines separating two
surface TRIM.  They do not, however, specify which of the TRIM are
{\it inside} of the surface Fermi surface and which are
{\it outside}.  We will show that the bulk parity invariants
$\delta(\Gamma_i)$ provide that information.

Specifically, for each surface TRIM
we will define the {\it surface fermion parity} as
the parity of the number of Fermi lines that enclose that TRIM.  This distinguishes
the TRIM that are outside the Fermi surface from those that are inside a (single)
electron or hole pocket.  For a crystal terminated on an inversion plane we
will establish a theorem which relates the
surface fermion parity to the bulk parity invariants.
Thus, for inversion symmetric crystals, the 8 bulk parity invariants
provide {\it more} information about the surface states than just the four
$\mathbb{Z}_2$ invariants.  We will give a simple
proof of this theorem in appendix A, which establishes a more direct
connection between the bulk parity eigenvalues and the surface state
structure than that presented in Ref. \onlinecite{fukane2}.

For the 111 surface of Bi$_{1-x}$Sb$_x$ our general theorem is consistent with
both our surface state calculation and with experiment.  We will also apply this
result to make predictions about the other surfaces of Bi$_{1-x}$Sb$_x$.
In addition, our theorem has implications
for inversion symmetric crystals which are ordinary insulators.
In particular, we will show that it has non trivial implications for the surface
states of pure Bi, whose valence band is in the trivial $(0;000)$
topological class.

(3) In addition to inversion symmetry, the crystal lattices of Bi and
Sb have a mirror symmetry.  We will show that the presence of mirror
symmetry leads to a further topological
classification of the bulk band structure in terms of
an integer, $n_{\cal M}$, which we refer to as a {\it mirror Chern number}.
This integer is similar to the spin Chern number, which occurs
in the quantum spin Hall effect when spin is conserved\cite{sheng}, and its parity
is related to the $\mathbb{Z}_2$ invariant\cite{fukane1}.
The valence band of pure Bi, which has the (0;000) $\mathbb{Z}_2$ class\cite{fukane2},
has $n_{\cal M}=0$.   The semiconducting alloy is a topological insulator with $\mathbb{Z}_2$ class
$(1;111)$.  There are two possibilities for the mirror Chern number $n_{\cal M}=\pm 1$, however,
which correspond to topologically distinct phases.
We will show that the sign of $n_{\cal M}$ in the topological insulator phase
further constrains the behavior of the surface states.

The transition between the $(0;000)$ and $(1;111)$ classes in Bi$_{1-x}$Sb$_x$ occurs
for small $x\sim .03$ because pure Bi is very close to a band inversion transition
where the $L_s$ valence band and $L_a$ conduction band cross.
The ${\bf k}\cdot {\bf p}$ theory of these states has been studied extensively in the literature
\cite{cohen,wolff,baraff,buot}
and has the form of a nearly massless {\it three dimensional} Dirac point.
We will show that the {\it change} $\Delta n_{\cal M}$ in the mirror Chern number at the band
inversion transition is determined by
a previously unexplored parameter in that theory: a sign $\eta = \pm 1$
 which we will refer to as the {\it mirror chirality}.
$\eta$ is related to the sign of the $g$ factor, which relates the magnetic moment to
the angular momentum in a particular direction.  For $\eta = +1$ the $g$ factor is like
that of a free electron, while for $\eta = -1$ it is anomalous.

We will use this result to interpret our surface state calculation
and to provide guidance for how $\eta$ can be measured.
In addition to the Dirac point enclosed by the surface Fermi surface
at $\bar\Gamma$, our tight binding surface band calculations for both pure Bi and
Bi$_{1-x}$Sb$_x$ predict that the 6 hole pockets also
enclose Dirac points which reside at points along the line between
$\bar \Gamma$ and $\bar M$.  Unlike the Dirac points at the surface TRIM, the
degeneracy at these Dirac points is not protected by time reversal
symmetry, but rather by  mirror symmetry.
This prediction is inconsistent with first principles calculations of the surface
states in Bi\cite{hirahara2,koroteev2}, which do not find a band
crossing inside the hole pocket.  Since the
Dirac point occurs above the Fermi energy ARPES experiments do not directly probe this
issue.  Nonetheless, spin resolved ARPES experiments on Bi
provide evidence that the surface
band structure is consistent with the first principles calculations\cite{hirahara2}.

We will show that this inconsistency can be traced to the mirror
chirality and the mirror Chern number.
The mirror chirality in the topological insulator phase
of Bi$_{1-x}$Sb$_x$ can be determined from the
structure of the ${\bf k}\cdot{\bf p}$ perturbation theory of the energy bands
in the vicinity of the $L$ point in pure Bi.
We find the Liu-Allen model predicts that $n_{\cal M} = +1$.  This value implies that the surface
state bands in the alloy cross in such a way as to establish the presence of the Dirac points in
the hole pockets in agreement with our surface state calculation.
In contrast, we find that an earlier, but more fundamental pseudopotential calculation by
Golin\cite{golin} predicts that $n_{\cal M}=-1$.  This value predicts that the bands do not cross,
and that there are
no extra Dirac points, which is consistent with the presently available experimental results
as well as first principles
calculations\cite{koroteev2,hirahara2}.
The Liu Allen tight binding parameters were chosen to reproduce the {\it energy}
of the bands computed using first
principles calculations, incorporating available experimental constraints.
Therefore, there is no reason to expect that it gets $n_{\cal M}$ right.  We
conclude that the inconsistency in our surface state calculation is
an artifact of the Liu Allen tight binding model, which culd be corrected
with a suitable choice of new parameters.

The outline of the paper is as follows.  In section II we will review the salient features of bulk
Bi$_{1-x}$ Sb$_x$ and describe our phenomenological tight binding model. In section III we will
describe our surface state calculations for Bi$_{1-x}$ Sb$_x$.   In section IV we will establish
the relationship between the surface fermion parity and the bulk parity eigenvalues and use that
result to analyze the surfaces of Bi$_{1-x}$ Sb$_x$.
In section V we will discuss the mirror Chern
number, and show it is related to the mirror chirality of the ${\bf k}\cdot{\bf p}$ theory
of pure Bi.  In section VI we will conclude with a discussion of the
relevance of our results to existing and future experiments.  Finally, in appendix A we provide a
simple proof of the theorem relating the surface fermion parity to the bulk parity eigenvalues.

\section{Bulk Bi$_{1-x}$ Sb$_x$}

\subsection{Introduction}

\begin{figure}
\centerline{ \epsfig{figure=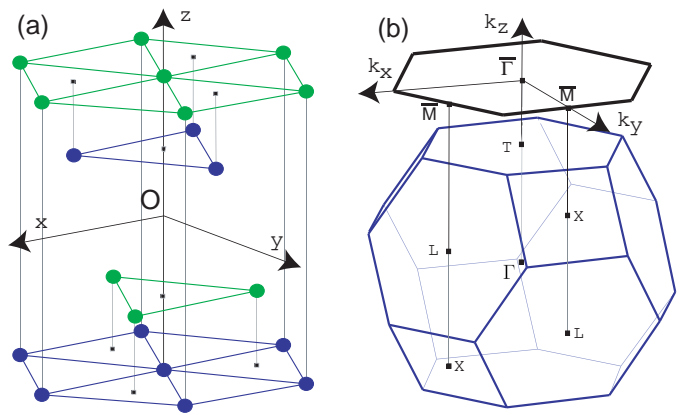,width=3.5in} }
 \caption{(a) Crystal structure of Bi. (b) 3D Brillouin zone and its projection onto the (111)
surface. Also displayed is the choice of coordinate system throughout the paper: z is along the
(111) direction, y is along the $\bar{\Gamma}$ to $\bar{M}$ direction, and $O$ is a center of inversion.
 }
 \label{crystal}
\end{figure}

Bismuth and Antimony are group V semimetals. They have the rhombohedral A7 structure shown
in Fig. \ref{crystal}(a), which
can be viewed as a distorted simple cubic lattice in which the triangular
(111) lattice planes (which we will refer to as monolayers) are paired to form bilayers.
The trivalent $s^2p^3$ atoms tend
to form strong covalent bonds directed to the three nearest neighbors within a bilayer.  Different
bilayers are more weakly coupled.
The primitive unit cell consists of two atoms in different monolayers, and each bilayer has a
structure similar to a honeycomb lattice.
The Brillouin zone for this lattice is shown in Fig. \ref{crystal}(b).
It contains 8 special points which are
invariant under inversion and time reversal,
denoted by $\Gamma$, $T$ and 3 equivalent $L$ and $X$
points.

Both Bi and Sb have a finite direct energy gap throughout the Brillouin zone, but they have a
negative indirect gap.  In Bi the conduction band minimum at $L$ is below the valence band
maximum at $T$, which gives rise
to an anisotropic hole pocket and three electron pockets with small effective masses\cite{smith}.
At $L$ the conduction band minimum, which has even parity $L_s$ symmetry, nearly touches
the valence band maximum, with odd parity $L_a$ symmetry, forming a three dimensional Dirac point
with a small mass gap $E_g \approx 11$ meV.   In Sb, the conduction band minimum at $L$
has $L_a$ symmetry, and is below the valence band maximum at the lower symmetry $H$ point.

The alloy Bi$_{1-x}$ Sb$_x$ retains the rhombohedral A7 crystal structure.  The evolution of its
band structure has been
studied experimentally
\cite{lenoir1,lenoir2}.
As $x$ is increased from zero two things happen.  First, the small gap
at $L$ closes and then reopens. The $L_s$ and $L_a$
bands switch places, and the mass of the three dimensional Dirac point changes sign.  Second, the top
of the valence band at $T$ descends below the bottom of the conduction band, resulting
in a semimetal-semiconductor transition.  For $.09 < x< .18 $ the alloy is a direct gap
semiconductor a gap of order 30 meV at the L points.

\subsection{Topological Invariants}

Time reversal invariant band structures are classified topologically
by four $\mathbb{Z}_2$ invariants \cite{moore,roy1,fkm}.
In Ref. \onlinecite{fkm} we exploited inversion symmetry to show that these
four invariants can be determined by the parity
$\xi_m(\Gamma_i)$
of the occupied bands at the 8 TRIM $\Gamma_i$, via the quantities
\begin{equation}
\delta(\Gamma_i) = \prod_n \xi_{2n}(\Gamma_i),
\label{delta}
\end{equation}
which we will refer to as {\it parity invariants}.
Here the product includes each Kramers pair (which satisfy $\xi_{2n}=\xi_{2n-1}$) only once.
For an inversion symmetric crystal, all 8 of the parity invariants are topological invariants in
the following sense.  If the crystal Hamiltonian is smoothly deformed, {\it preserving the
inversion symmetry}, then the only way any of the $\delta(\Gamma_i)$'s can change is if the gap at
$\Gamma_i$ goes to zero, so that states with opposite parity can be exchanged between the
conduction and valence band.  If inversion symmetry is relaxed, then the 8 invariants lose their
meaning.  However, in Ref. \onlinecite{fukane2} we showed that provided
{\it time reversal symmetry} is preserved
four combinations of the $\delta(\Gamma_i)$ remain robust and define the four
$\mathbb{Z}_2$ invariants denoted by $(\nu_0;\nu_1\nu_2\nu_3)$.  The most
important invariant, $\nu_0$,  distinguishes the strong topological
insulator, and survives even in the presence of disorder\cite{fkm,fukane2}.
$(-1)^{\nu_0}$ is given simply by the product of all 8 $\delta(\Gamma_i)$.

Pure Bi and Sb have inversion symmetry.  The parity eigenvalues for inversion about the
point O in Fig. \ref{crystal}(a) are tabulated in the literature\cite{falikov,golin,liuallen}.
Based on this data
we display $\delta(\Gamma_i)$ in Table \ref{deltatab}, along with the predicted
$\mathbb{Z}_2$ invariants for pure Bi, pure antimony and the alloy.
The valence band of pure Bi is characterized by the trivial
class $(0;000)$, while antimony has the $(1;111)$ class.
The difference is due to the inversion of the
$L_s$ and $L_a$ bands, which changes the sign of $\delta(L)$.  The
alloy inherits its topological class from antimony, and is a strong
topological insulator.

\begin{table}
  \centering
  \begin{tabular}{|c|cccc|c|}
\hline
& $\delta(\Gamma)$ & $\delta(L)$ & $\delta(T)$ & $\delta(X)$  & $(\nu_0;\nu_1\nu_2\nu_3)$\\
\hline
Bismuth & -1 & -1 & -1 & -1 & (0;000) \\
 \hline
Antimony &  -1 & 1 & -1 & -1 & (1;111) \\
\hline
Bi$_{1-x}$Sb$_x$ &  -1 & 1 & -1 & -1 & (1;111) \\
\hline

\end{tabular}
  \caption{Parity invariants $\delta(\Gamma_i)$ and
  $\mathbb{Z}_2$ topological invariants $(\nu_0;\nu_1\nu_2\nu_3)$ for
 Bismuth, Antimony, and Bi$_{1-x}$Sb$_x$ determined from the product of parity eigenvalues
 $\xi_m(\Gamma_i)$ at each bulk TRIM $\Gamma_i$.}
  \label{deltatab}
\end{table}

\subsection{Pure Bi, Sb : Liu Allen model}

Liu and Allen\cite{liuallen} developed a third neighbor tight binding model for the electronic structure Bi and
Sb, which describes the atomic $s$ and $p$ orbitals nearest to the Fermi energy.  The Bloch
Hamiltonian $\hat H({\bf k}) = e^{-i{\bf k}\cdot{\bf r}} {\cal H} e^{i{\bf k}\cdot{\bf r}}$ has
the form
\begin{equation}
\hat H({\bf k}) =
\left(\begin{array}{cc} H_{11}({\bf k}) & H_{12}({\bf k}) \\ H_{21}({\bf k}) & H_{22}({\bf k}).
 \end{array}\right)
\label{hk}
\end{equation}
Here $H_{ab}({\bf k})$ are 8 by 8 matrices describing the coupling between the 2 s states and 6 p
states on the $a$ and $b$ sublattices of the crystal.  The explicit form of these matrices is
given in Tables IX and X in the appendix of Ref. \onlinecite{liuallen}.

$H_{11} = H_{22}$ describe the coupling within the same sublattice.  These terms involve the
on site energies $E_s$ and $E_p$ as well an on site spin orbit coupling, $\lambda$.  The closest
neighbor on the same sublattice is the third neighbor, which resides in the same monolayer as the
origin.  The third neighbor hopping involves four parameters $V''_c$ with
$c = ss$, $sp\sigma$, $pp\sigma$ and $pp\pi$, describing
hopping between the $s$ and $p$ states.  Since further neighbor hopping is not included in this
model, $H_{11}({\bf k})$ and $H_{22}({\bf k})$ describe decoupled monolayers and depend only
on the momentum ${\bf q} = {\bf k}_\parallel$ in the plane of the monolayer.

$H_{12} = H_{21}^\dagger$ describes the coupling between the sublattices.  These involve two terms :
First neighbor hopping terms $V_c$ couples atoms within the same bilayer, and second neighbor
hopping terms $V'_c$ couple atoms in neighboring bilayers.  In the following it will be useful to
separate these two contributions by writing ${\bf k} = ({\bf q},k_z)$
\begin{equation}
H_{12}({\bf q},k_z) = H_{12}^{(1)}({\bf q}) e^{ik_z c_1} +
H_{12}^{(2)}({\bf q}) e^{-ik_z c_2},
\label{h12}
\end{equation}
where $c_1$ and $c_2$ are the spacing between the monolayers within a bilayer and between different
bilayers, and ${\bf q}$ and $k_z$ are the momenta parallel and perpendicular to the surface.
$H_{12}^{(1)}$ and $H_{12}^{(2)}$ can be extracted from Table X of Ref. \onlinecite{liuallen} by noting
that they are the terms which involve the parameters $g_0-g_{12}$ and $g_{13}-g_{26}$
respectively.

The 12 hopping parameters and 3 on site parameters make a total of 15 parameters specifying this
model.  These were chosen to reproduce the
energies predicted by first principles calculations, as well as details of the band gaps and
effective mass tensors which are known experimentally.  The values of the parameters for both
Bi and Sb are listed in table II of Ref. \onlinecite{liuallen}.

\subsection{Tight binding model for alloy}

In order to describe the electronic structure of the alloy Bi$_{1-x}$Sb$_x$, we wish to develop a
``virtual crystal" approximation which treats the substitutional disorder in mean field theory and
results in a translationally invariant effective Hamiltonian.  Since the regime of interest is $x
\sim 0.1$, the effective Hamiltonian should be close to that of pure Bi.  The effect of small $x$
will be to modify the band energies, but not drastically change the wavefunctions.   The effective
Hamiltonian should
reproduce two essential features:  (1) the inversion of the $L_s$ and $L_a$ bands (which are
nearly degenerate in pure Bi), and (2) the
descent of the valence band at T below the conduction band at L, as $x$ is increased, which leads
to the transition between the semimetal and the semiconductor.

\begin{figure}
\centerline{ \epsfig{figure=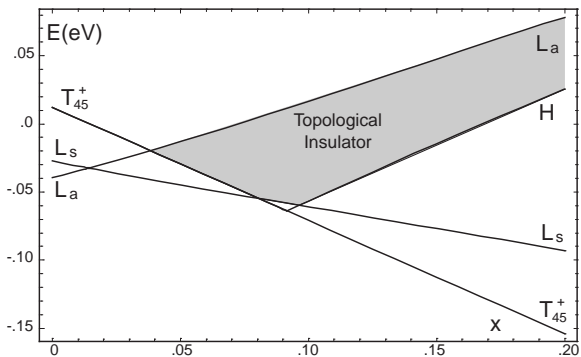,width=3in} }
 \caption{Band evolution of interpolated tight binding model using
 the parameters in (\ref{alphac},\ref{vc})
 }
 \label{bandevolution}
\end{figure}

The simplest approach would be to simply interpolate between the tight binding parameters for
Bismuth and Antimony.  For each of the 15 tight binding parameters $\alpha_c$, we could define
\begin{equation}
\alpha_c(x) = x \alpha_c^{\rm Sb} + (1-x) \alpha_c^{\rm Bi}.
\label{alphac}
\end{equation}
However, for this simple interpolation the inversion between $L_s$ and $L_a$ occurs at a rather
large value $x\sim .4$, which occurs after the semimetal-semiconductor transition.
We found that this could be corrected if each of the hopping terms
(but not the other terms) are revised such that
\begin{equation}
V_c(x) = x V_c^{\rm Sb} + (1-x^2) V_c^{\rm Bi}.
\label{vc}
\end{equation}
This approach is admittedly ad hoc, but it is sufficient for our purposes because it correctly
accounts for the most important features of the band evolution.
In Fig. \ref{bandevolution} we plot the energies of the
$T_{45}^-$, $L_s$ and $L_a$ as a function of $x$ for this model.  The qualitative behavior of the
known band evolution is reproduced, including the decent of the hole pocket at $T$ and the
inversion of the conduction and valence bands at $L$.  This should not, however, be interpreted as
a quantitative description of the band evolution of Bi$_{1-x}$ Sb$_x$.

\section{Surface states of  Bi$_{1-x}$ Sb$_x$}

In this section we describe our calculation of the 111 surface band structure for a semi infinite
lattice in the half plane $z<0$ described by the interpolated tight binding model described above.
We begin with a brief
discussion of our method, which is based on a transfer matrix scheme\cite{dhlee},
and then go on to discuss the results.

\subsection{Transfer matrix method}

The electronic states of a semi infinite crystal can be represented as
$\phi_{n, a}({\bf q})$ in a basis of states
which are plane waves with momentum ${\bf q}$ in the plane of the
surface, but are localized on the $a=1,2$ monolayer of the $n$th bilayer.  Each $\phi_{n a}$ has
8 components associated with the 8 atomic orbitals.
 The time independent Schrodinger equation, written in this basis may be expressed in the form
\begin{equation}
\left(\begin{array}{c} \phi_{n+1,1} \\ \phi_{n+1,2} \end{array}\right) = T({\bf q},E)
\left(\begin{array}{c} \phi_{n,1} \\ \phi_{n,2} \end{array}\right),
\end{equation}
where the transfer matrix is given by $T({\bf q},E) = t_{11}({\bf q},E) t_{22}({\bf q},E) $, with
\begin{equation}
t_{11}=\left(\begin{array}{cc}{H_{21}^{(2)}}^{-1}(E-H_{22}) &
-{H_{21}^{(2)}}^{-1}H_{21}^{(1)} \\
1 & 0\end{array}\right),
\end{equation}
and
\begin{equation}
t_{22}=\left(\begin{array}{*{20}c}{H_{12}^{(1)}}^{-1}(E-H_{11})
& -{H_{12}^{(1)}}^{-1}H_{12}^{(2)} \\ 1 & 0\end{array}\right).
\end{equation}
Any bulk state is an eigenstate
of the $16\times 16$ transfer matrix with unimodular eigenvalues. For $E$ within the energy gap,
$T({\bf q},E)$ has
exactly eight eigenvalues with modulus larger than $1$.  These correspond to states that
decay exponentially in the $-z$ direction.
$E({\bf q)}$ will correspond to a surface state localized at the top
surface in Fig. \ref{tbsurface}(a)
near $z=0$ provided there is a linear
combination of the decaying states which vanish on the monolayer $n=0$, $a=1$
just outside the surface:
$\phi_{0,1}=0$.   The surface states are thus determined by forming an 8 by 8 matrix $M({\bf q},E)$
composed of the 8 components of $\phi_{0,1}$ for each of the 8 decaying states.  $E({\bf q})$ is
then determined by solving ${\rm det}[M({\bf q},E)]=0$.

\subsection{Electronic structure of (111) surface}

\begin{figure}
\centerline{ \epsfig{figure=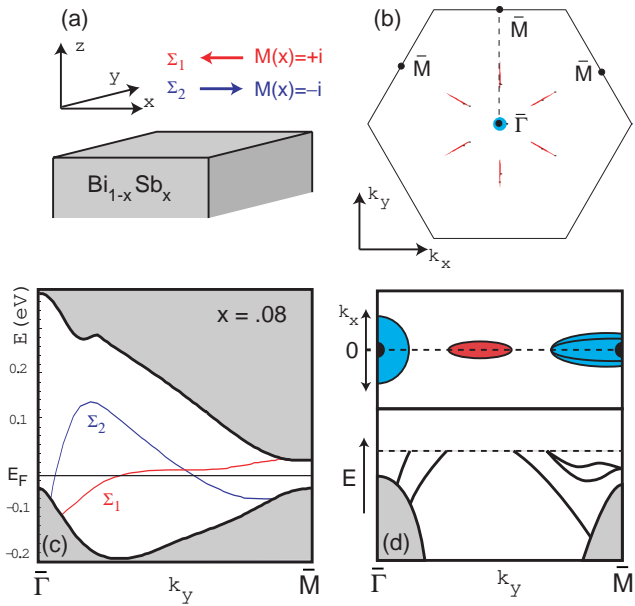,width=3.3in} }
\caption{(a) Geometry for our surface state calculations, which defines our coordinate system and
specifies the spin directions of the $\bar\Sigma_1$ and $\bar\Sigma_2$ bands, which have mirror
eigenvalues $+i$ and $-i$ respectively.  (b) Brillouin zone
for the (111) face of Bi$_{1-x}$Sb$_x$ with the electron pocket and six hole pockets predicted by our tight
binding calculation.  (c) Surface band structure along the line between $\bar\Gamma$ and $\bar M$
predicted by tight binding model.  The shaded regions are the bulk states projected to the surface.
(d) Schematic
illustration of experimental surface band structure and Fermi surface probed by angle resolved
photoemission spectroscopy\cite{hasan}. The top shows the Fermi surface in a slice of the Brillouin zone
near $k_x=0$, and the bottom shows the surface state dispersion.
Compared with (c), there are two additional bands near $\bar{M}$.
}
 \label{tbsurface}
\end{figure}

Fig. \ref{tbsurface}(c) shows the energy spectrum of the (111) surface states of Bi$_{1-x}$Sb$_x$ for
$x = .08$ calculated along the line connecting ${\bf q}=\bar\Gamma=0$
to $\bar M$ along the $+\hat y$ axis using the transfer matrix method for the interpolated tight binding
model.  Fig. \ref{tbsurface}(b) shows the Fermi surface.  We find two bound surface states within the
bulk energy gap.  Along the line $q_x=0$ these states are labeled by their symmetry under
the mirror ${\cal M}(\hat x)$ which takes $x$ to $-x$.  Since the mirror
operation also operates on the spin degree of freedom it is important to be specific
about its definition.  We write
$M(\hat x)=P C_2(\hat x)$,
where $P$ is inversion, and $C_2(\hat x)$ is a $180^\circ$ counterclockwise
rotation about the positive $\hat x$ axis.  $P$ does not affect the spin degree of freedom,
but the $C_2$ rotation does.  The resulting
eigenvalues of $M(\hat x)$ are $+i$ and $-i$, which we label as $\bar\Sigma_1$
and $\bar\Sigma_2$.
These mirror eigenvalues are correlated with the spin $S_x$.
For a free spin, eigenstates with $M(\hat x) = \pm i$ correspond to spin eigenstates with $S_x = \mp
\hbar/2$.  The surface states are not spin eigenstates, but on the line
$k_x=0$, $0<k_y < k_y(\bar M)$ the expectation value of the spin
satisfies $ \langle \vec S \rangle \propto i \langle M(\hat x) \rangle \hat x \propto -(+)\hat x$,
for $\Sigma_{1(2)}$, as indicated in  Fig. \ref{tbsurface}(a).

The Fermi surface shown in Fig. \ref{tbsurface}(b) consists of electron and hole pockets.
A single electron pocket surrounds $\bar
\Gamma$.  This Fermi surface is non degenerate, and opposite sides of the Fermi surface are Kramers
pairs with opposite spin.  The electronic states pick up a Berry's phase of $\pi$
when they are adiabatically transported around the Fermi surface.  This can be understood to be a
consequence of the $360^\circ$ rotation of the spin going around the Fermi surface.  The Fermi
surface is thus spin filtered, in the sense that the spin of the electron is correlated with
it's propagation direction, roughly satisfying $\langle \vec S \rangle \propto \hat q \times \hat
z$ for an electron propagating in the $\hat q$ direction in the plane.
In addition, there are 6 elliptical hole pockets centered along the 6 lines connecting
$\bar\Gamma$ to $\bar M$.  These are also non degenerate, though unlike the electron pocket, the
time reverse of a hole pocket is a different hole pocket.  The crossing of the $\bar\Sigma_1$ and
$\bar\Sigma_2$ bands is protected by the mirror symmetry for $q_x=0$.  The degeneracy will be lifted for
finite $q_x$, so the crossing describes a two dimensional {\it Dirac} point, which is enclosed by the
hole pocket.

\subsection{Comparison with topological predictions}

A single band of surface states connects the valence and conduction band between $\bar\Gamma$ and
$\bar M$ in Fig. \ref{tbsurface}(d).
This confirms the topological predictions for the connectivity of the surface state
bands.  In Ref. \onlinecite{fkm} we showed that the number of times $\Delta N(\Lambda_a,\Lambda_b)$
the surface states intersect the Fermi
energy between two surface TRIM $\Lambda_a$ and $\Lambda_b$ satisfies
\begin{equation}
(-1)^{\Delta N(\Lambda_a,\Lambda_b)} = \pi(\Lambda_a)\pi(\Lambda_b),
\label{deltan}
\end{equation}
where
\begin{equation}
\pi(\Lambda_a) = (-1)^{n_b} \delta(\Gamma_{a1})\delta(\Gamma_{a2}).
\label{pi}
\end{equation}
Here $\Gamma_{a1}$ and
$\Gamma_{a2}$ are the two bulk TRIM which project to the surface TRIM $\Lambda_a$.
The 8 parity invariants $\delta(\Gamma_i)$, defined in Eq. \ref{delta}, are products of
parity eigenvalues.
This definition of $\pi(\Lambda_a)$ differs slightly from the one introduced in Refs. \onlinecite{fkm,fukane2}
 because of
the additional factor $(-1)^{n_b}$.  $n_b$ is the number of occupied Kramers degenerate pairs of energy
bands, which is equal to the number of terms in the product of Eq. \ref{delta}.
For Bi$_{1-x}$Sb$_x$, $n_b=5$.  This factor does not affect
$\Delta N(\Lambda_a,\Lambda_b)$ in (\ref{deltan}).  However, this
modification simplifies our further results, discussed below.

For $\pi(\Lambda_a) \pi(\Lambda_b)= -1$ there will be an odd number of crossings between
$\Lambda_a$ and $\Lambda_b$, guaranteeing
the presence of the gapless surface states.
In the appendix we will provide a new derivation of this connection between the surface states and the
bulk parity eigenvalues which is simpler and more direct than our previous proof\cite{fukane2}.
This will show that with inversion symmetry the
eight parity invariants $\delta(\Gamma_i)$ contain more information about the surface
state structure than just the
number of crossings, a fact we will exploit in section $IV$ to make general predictions about the
locations of electron and hole pockets in the surface Brillouin zone.

From Fig. \ref{crystal}, Table \ref{deltatab} and Eq. \ref{pi} it can be seen that for the alloy,
\begin{eqnarray}
&\pi(\bar\Gamma)  = -\delta(\Gamma)\delta(T) = -1, \\
&\pi(\bar M)  = -\delta(X)\delta(L) = +1.
\end{eqnarray}
This predicts that there should be an odd number of crossings between $\bar\Gamma$ and $\bar M$,
which is confirmed both by our explicit calculation and, as we will discuss below, by experiment.

\subsection{Comparison with experiment}

Before comparing our calculation to experiment and other calculations, it is worthwhile to discuss
what our calculation does {\it not} include.  In addition to our approximate treatment of the
alloy's bulk electronic structure, we have made no attempt to self consistently describe the
potential near the surface.  This will be modified by relaxation of the bonds near the surface.
More importantly, the population of the surface states determines the electric charge distribution
near the surface, which leads to Hartree and exchange contributions to the potential. We assume
that the surface is electrically neutral.  We will argue in section III that this means that the
area of the electron pocket is equal to the total area of the six hole pockets.  However the
potential due to a surface dipole layer is not included in our calculation. The effect of such a
surface potential will be to modify the energies of the bands and perhaps to split off additional
surface state bands from the continuum.  However, the topological connectivity of the surface
state bands will not be altered.

In their recent ARPES experiment\cite{hasan}, Hsieh et al. measured the spectrum of Bi$_{.9}$ Sb$_{.1}$ (111)
surface states below $E_F$ between $\bar\Gamma$ and $\bar M$.  The observed spectrum,
which we have sketched schematically in Fig. \ref{tbsurface}(d), resembles
Fig. \ref{tbsurface}(c), though there are some important differences.  As in Fig.
\ref{tbsurface}(c), two surface state bands
emerge from the bulk valence band near $\bar\Gamma$.  The first intersects the Fermi energy
forming the electron pocket centered on $\bar\Gamma$, while the second intersects the Fermi energy
forming a hole pocket.  A third band crosses $E_F$ from above, forming the opposite side of the
hole pocket, and merges with the bulk valence band near $\bar M$.  Unlike our calculation, the
observed spectrum includes an additional electron pocket near $\bar M$.  A Kramers degenerate pair
of surface states is found in the gap at $\bar M$. Away from $\bar M$ these states split to form
two surface bands, which both cross $E_F$ near the end of the hole pocket.  Thus there are a total
of five bands crossing $E_F$ between $\bar\Gamma$ and $\bar M$, which is consistent with the
prediction for a $(1;111)$ topological insulator.  The discrepancy between our calculation and the
experiment is most likely a consequence of our neglect of the self consistent surface potential,
which could lead to a Kramers pair of bound states to be split off from the conduction band at
$\bar M$.

\begin{figure} \centerline{ \epsfig{figure=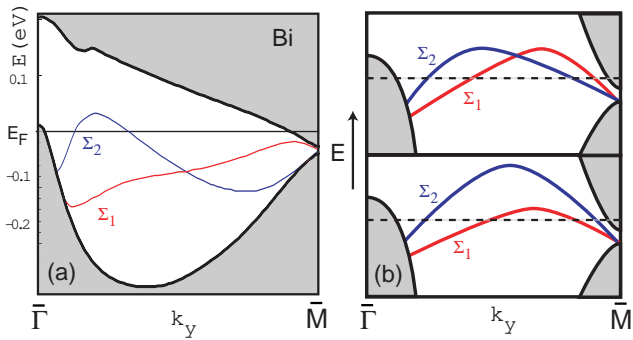,width=3.3in} }
\caption{(a) Bi surface states between $\bar\Gamma$ and $\bar M$
calculated using tight binding model.  (b) Schematic picture of Bi
bands in (a) in which Hartree effects raise the bands to accommodate
charge neutrality. The crossing of $\Sigma_1$ and $\Sigma_2$ results
in a Dirac point enclosed by a hole pocket. (c) Schematic picture
without the crossing between $\Sigma_1$ and $\Sigma_2$, which
resembles a first-principle calculation of surface states in
Bi\cite{koroteev2,hirahara2}}
 \label{tbsurfacebi}
\end{figure}

It is also instructive to compare our calculation with previous experimental and theoretical
results for pure Bi.  In Fig. \ref{tbsurfacebi}(a) we show the surface state
spectrum for pure Bi
calculated using the transfer matrix method for the Liu-Allen tight binding model.  The number of
band crossings is consistent with the trivial $(0;000)$ topological structure of the
Bi valence band.  Since the
Fermi energy of semimetallic Bi is fixed by the bulk, our calculated surface states violate surface
charge neutrality: too many surface states are occupied, so the surface will
have a negative charge.  Hartree effects will push the surface states up in energy, but they will not
alter the topological connectivity of the surface states.  This allows us deduce qualitative
conclusions from the calculation.

First, as in our alloy calculation, two surface bands emerge from the bulk valence band near $\bar
\Gamma$.  These are also seen in photoemission experiments as well as first principles
calculations on pure Bi\cite{koroteev2}.
Moreover, the spin $\langle S_x \rangle$ of those surface states
has been both calculated and measured using spin polarized ARPES\cite{hirahara2}.
We have checked that the spin direction
predicted by our tight binding calculation for each of these bands agrees with the experimental
and first principles theory results.  Thus, the behaviour near $\bar\Gamma$, including the ordering
in which the $\bar\Sigma_2$ emerges first and forms the electron pocket
appears to be robust, with all calculations in agreement with each other and with experiment.

There is a discrepancy, however, between the tight binding calculation and the first principles
calculation\cite{koroteev2,hirahara2}. The crossing between the $\bar\Sigma_1$ and $\bar\Sigma_2$ bands in Figs.
\ref{tbsurface}(c) and \ref{tbsurfacebi}(a) is not
found in the first principles calculation.  Since it is likely that this crossing would be pushed
above the Fermi energy by Hartree corrections (so that the crossing occurs inside a hole pocket),
the tight binding model predicts that the hole pockets of Bi (111) enclose a Dirac point,
as shown schematically in Fig. \ref{tbsurfacebi}(b).
The existence of this band crossing is not directly
probed by ARPES which only probes occupied states, though it could be
probed using inverse photoemission.  There is, however, {\it indirect} experimental evidence that the
crossing does {\it not} occur.  Spin polarized ARPES measurements\cite{hirahara2} have
measured the spin on both
sides of the hole pocket.  Though the signal appears weak, the sign of the spin is resolved, and
determined to be
the same on both sides, indicating that there is no crossing, as shown schematically
in Fig. \ref{tbsurfacebi}(c).  This agrees with the predictions of the first principles
calculations that both
sides are in the same $\bar\Sigma_1$ band.  In contrast, our tight binding model predicts
that the opposite sides of the hole pockets correspond to the $\bar\Sigma_1$ and $\bar\Sigma_2$
bands, which have opposite spin.

It thus appears likely that the prediction of the level crossing which
implies that the hole pockets enclose a Dirac point is an
artifact of the tight binding model.  This brings into question the related prediction of the tight
binding model that the hole pockets of the alloy also enclose a Dirac point.
In Section V we will argue that this artifact is a consequence of a subtle error in the
Liu-Allen tight binding model.

\section{Inversion symmetry and the surface fermion parity}

An inversion symmetric crystal can have no bulk electric
polarization.  In this section we show that this fact in combination
with surface charge neutrality has non trivial implications for the
surface state structure because it allows the {\it outside} of the surface Fermi
surface to be unambiguously defined.  It is then possible to define electron
pockets to be regions in the surface Brillouin zone
where an extra band is occupied and hole pockets as regions where an otherwise
occupied band is empty.  Charge neutrality dictates that the area of the electron pockets
should equal that of the hole pockets.  We will show that the locations of the
electron and hole pockets in the surface Brillouin zone are topologically constrained by the bulk
parity invariants $\delta(\Gamma_i)$.  In addition to fixing the number of Fermi energy crossings,
we find
that $\delta(\Gamma_i)$ determine which TRIM are on the {\it inside} of an electron or
hole pocket and which TRIM are on the {\it outside}.  We define the {\it surface
fermion parity}, which specifies whether a given surface TRIM is enclosed by an even or odd number of
Fermi lines.  We will begin with a general discussion of the relationship
between the surface fermion parity to the bulk parity invariants.  We will then apply our general
result to the surfaces of Bi$_{1-x}$Sb$_x$ and Bi.

\subsection{Surface fermion Parity}

The total surface charge density may be expressed as a sum over the surface
Brillouin zone (SBZ),
\begin{equation}
\sigma = e \int_{SBZ} {d^2 q\over{(2\pi)^2}} N({\bf q}),
\label{surfacecharge}
\end{equation}
where the surface fermion number $N({\bf q})$ represents the excess charge in the vicinity of the
surface due to states with momentum ${\bf q}$ in the plane of the surface.
If we assume that the bulk Fermi energy is inside the gap, then there will be two contributions,
$N({\bf q}) = N_{\rm bulk}({\bf q}) + N_{\rm surface}({\bf q})$.
$N_{\rm surface}({\bf q})$ is an integer which counts the occupied discrete surface states inside the
energy gap. $N_{\rm bulk}({\bf q})$
is the total surface charge in the continuum valence band states.
For a crystal with inversion symmetry there can be no bulk electric
polarization, and $N_{\rm bulk}({\bf q})$ will
also be quantized.   In Appendix A we will show that it must be an integer\cite{claro}.

The integer values of $N({\bf q})$ allow us to unambiguously define the ``outside" of the surface
Fermi surface to be the region for which $N({\bf q}) = 0$.  $N({\bf q})= +(-) 1$ define electron
(hole) pockets.  $N({\bf q}) = +(-) 2$ is a double electron (hole) pocket, and so on.  From
(\ref{surfacecharge}), charge
neutrality implies that the total area of the electron pockets equals that of the hole pocket, provide
the double pockets are appropriately counted.

Kramers' theorem requires that the surface states be two fold degenerate at the TRIM ${\bf q} =
\Lambda_a$ in the surface Brillouin zone.  Provided the Fermi energy is not exactly at the
degeneracy point this means that $N_{\rm surface}(\Lambda_a)$ is even, so that the {\it parity} of
$N(\Lambda_a)$ is equal to the parity of $N_{\rm bulk}(\Lambda_a)$.  In Appendix A we will show
that the {\it surface fermion parity} is determined by the bulk
parity invariants,
\begin{equation}
(-1)^{N(\Lambda_a)} \equiv \pi(\Lambda_a) = (-1)^{n_b}
\delta(\Gamma_{a1})\delta(\Gamma_{a2}).
\label{fermionparity}
\end{equation}
Eq. \ref{fermionparity} determines whether the TRIM $\Lambda_a$ is enclosed by a single
(or odd number) of
Fermi lines, or whether it is outside the Fermi surface (or enclosed by an even number).
In the special case
that the Fermi energy is exactly at a Dirac point at $\Lambda_a$, $\Lambda_a$ should be
interpreted to be inside an electron (or hole) pocket with vanishing size.

Eq. \ref{fermionparity} is a new result which provides information about the structure
of the surface Fermi surface beyond that determined by the $\mathbb{Z}_2$ invariants
$(\nu_0;\nu_1\nu_2\nu_3)$.  We will show below that this result can have non trivial consequences
even in materials which are {\it not} topological insulators.  For example, we will see that
(\ref{fermionparity}) constrains the surface states of pure Bi.

\begin{figure}
\centerline{ \epsfig{figure=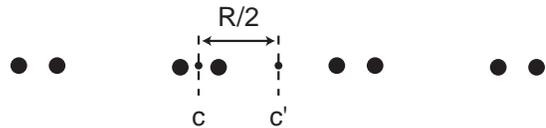,width=2.8in} }
\caption{Two inequivalent inversion centers $c$ and $c'$ in an inversion-symmetric crystal, which differ by
half a lattice vector. The parity eigenvalues of Bloch state at momentum $k=\pi/R$ with inversion
center chosen at $c$ and $c'$ are different. Crystals terminated at $c$ and $c'$ will have surface
charges that differ by an odd integer.
}
 \label{inversion}
\end{figure}

In order to apply (\ref{fermionparity}) it is essential to use the parity eigenvalues
associated with an inversion center in the plane on which the crystal
is terminated.  As a simple example Fig. \ref{inversion} shows a one
dimensional inversion symmetric lattice, which has two distinct
inversion points.
In general, a three dimensional inversion symmetric crystal has 8
distinct inversion centers, which are related to each other by {\it
half} a Bravais lattice vector: ${\bf c}' = {\bf c} + {\bf R}/2$.
The parity eigenvalues associated with inversion center ${\bf c}'$
will be related to those associated with ${\bf c}$ by
\begin{equation}
\xi_m'(\Gamma_i) = \xi_m(\Gamma_i) e^{i\Gamma_i\cdot{\bf
R}} = \pm \xi_m(\Gamma_i).
\end{equation}

An inversion plane will contain four of those points.  For a given
surface orientation there are two distinct parallel inversion planes.
For a surface terminated on one of those inversion planes,
$\pi(\Lambda_a)$ does not depend on which of the four inversion centers
within the inversion plane are used.  This can be seen by noting that
\begin{equation}
\pi'(\Lambda_a)=\pi(\Lambda_a)\exp[i n_b (\Gamma_{a1}-\Gamma_{a2})\cdot{\bf R}],
\label{piprime}
\end{equation}
where $n_b$ is the number of occupied bands.  When ${\bf c}$ and ${\bf c}'$ are in the plane of the surface
the dot product in the exponent is zero.
Crystals terminated on inequivalent inversion planes, however will have different
$N(\Lambda_a)$.  For odd $n_b$, $\pi'(\Lambda_a) = -\pi(\Lambda_a)$, so that
the parity of $N(\Lambda_a)$ changes at all four $\Lambda_a$.  Thus, changing the inversion plane
amounts to filling (or emptying) a single surface band throughout the surface Brillouin zone.
Since $N(\Lambda_a)$ depends on how the crystal is terminated,
it is not a bulk property.  However, $\Delta N(\Lambda_a,\Lambda_b) =
N(\Lambda_a)-N(\Lambda_b)\ {\rm mod}\ 2$  is a bulk property, which
is determined by the $\mathbb{Z}_2$ invariants $(\nu_0;\nu_1\nu_2\nu_3)$.

\subsection{Application to Bi$_{1-x}$Sb$_x$}

We now apply our general result to Bi$_{1-x}$Sb$_x$ surfaces.
In order to apply (\ref{fermionparity})
it is necessary to identify the appropriate inversion centers.  The 8
inversion centers of the rhombohedral A7 lattice are (1) ${\bf c}_0 = 0$,
the origin in Fig. 1, which is between two bilayers. (2-4)
${\bf c}_{j=1,2,3} = {\bf a}_j/2$.  Here ${\bf a}_j$
are the three rhombohedral primitive Bravais lattice
vectors, which connect an atom to the nearest three atoms on the
same sublattice of the neighboring bilayer\cite{review}.  These points are at the
center of a nearest neighbor bond
in the middle of a bilayer.  (5-7) ${\bf c}_{ij} \equiv ({\bf a}_i + {\bf a}_j)/2$ for
$i\ne j$.  These three points are at the center of a 2nd neighbor bond
between two bilayers.  (8) ${\bf c}_{123} = ({\bf a}_1+{\bf a}_2+{\bf a}_3)/2$, which
is directly above the origin in Fig. \ref{crystal}, in the middle of a bilayer.
For a given surface orientation, these inversion centers are divided
into two groups of four, which reside in two possible cleavage planes.

\begin{table}
  \centering
  \begin{tabular}{|l| cc |c |c|c|}

\multicolumn{1}{c}{  Face} & \multicolumn{2}{c}{${\bf c}_j$} &
\multicolumn{1}{c}{$\Lambda_a = (\Gamma_{a1}\Gamma_{a2})$} &
\multicolumn{1}{c}{$\pi_{\rm BiSb}(\Lambda_a$)} &
\multicolumn{1}{c}{$\pi_{\rm Bi}(\Lambda_a$)}\\
\hline
$(111)$ &
${\bf c}_0$ & ${\bf c}_{12}$ &
$\bar\Gamma=(\Gamma T)$ & $-1$ & $-1$ \\
&${\bf c}_{13}$ & ${\bf c}_{23}$ & $3\bar M = (LX)$ & $+1$ & $-1$ \\
 \hline
$(111)'$ &
${\bf c}_1$ & ${\bf c}_2$ &
$\bar\Gamma=(\Gamma T)$ & $+1$ & $+1$ \\
&${\bf c}_3$ & ${\bf c}_{123}$ & $3\bar M = (LX)$ & $-1$ & $+1$ \\
\hline
$(110)$ &
${\bf c}_0$ & ${\bf c}_3$ &
$\bar\Gamma = (\Gamma X)$ & $-1$ & $-1$ \\
&${\bf c}_{12}$ & ${\bf c}_{123}$& $\bar X_1 = (LL)$ & $-1$ & $-1$ \\
&&& $\bar X_2 = (LT)$  & $+1$ & $-1$ \\
&&& $\bar M = (XX)$ & $-1$  & $-1$ \\
\hline
$(100)$ &
${\bf c}_1$ & ${\bf c}_{13}$ &
$\bar\Gamma = (\Gamma L)$ &  $-1$ & $+1$\\
&${\bf c}_{23}$ & ${\bf c}_{123}$ & $\bar M = (TX)$ & $+1$  & $+1$\\
&&& $2\bar M' = (LX)$ &  $-1$ & $+1$\\
\hline

\end{tabular}
  \caption{For each crystal face $(hkl)$ we list the 4 inversion centers $\bf c_j$ on the cleavage plane
  along with the projections relating the 4 surface TRIM ${\Lambda}_a$
  to the bulk TRIM $\Gamma_{a1,2}$.  For each ${\Lambda}$ we
list the surface fermion parity, $\pi(\Lambda_a)$ for both Bi$_{1-x}$Sb$_{x}$ and Bi.
$\pi(\Lambda_a)$  is a product of parity invariants at $\Gamma_{a1,2}$.
}
  \label{fptab}
\end{table}

In Ref. \onlinecite{review}, the (111), (110) and (100) faces of Bi are
discussed, where the Miller indices $(mno)$ refer to the rhombohedral
reciprocal lattice vector $m {\bf b}_1 + n{\bf b}_2 + o {\bf b}_3$ with
${\bf a}_i \cdot {\bf b}_j = 2\pi\delta_{ij}$.
In these cases the preferred cleavage plane is the
one which minimizes the number of broken first neighbor bonds.
In table \ref{fptab} we list the four inversion centers in the cleavage plane
for each of these faces.  For comparison, we have also included the
$(111)'$ face, which is terminated in the middle of a bilayer
(breaking three nearest neighbor bonds).  Table \ref{fptab} also
shows how the bulk TRIM project onto the surface TRIM, using the
notation $\Lambda_a = (\Gamma_{a1}\Gamma_{a2})$.  This data,
combined with Table \ref{fptab}
is sufficient to determine the surface fermion parity
$\pi(\Lambda_a)$ for both the alloy Bi$_{1-x}$Sb$_x$ (BiSb) and
pure Bi for each surface as shown in Table \ref{fptab}.

\begin{figure}
\centerline{ \epsfig{figure=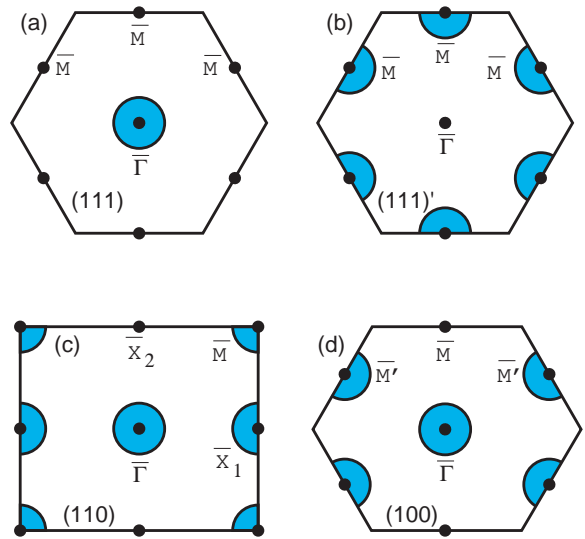,width=3.0in} }
\caption{ Schematic diagram showing which surface TRIM are enclosed by
an odd number of electron or hole pockets for different faces of
Bi$_{1-x}$Sb$_x$ predicted by the surface fermion parity in
Table \ref{fptab}.  (a-d) show the
(111), (111)', (110) and (100) faces.  The (111)' surface
is a hypothetical surface cleaved in the
middle of a bilayer.}
\label{bzfig}
\end{figure}

First consider
the 111 surface.  The parity eigenvalues
quoted in the literature, which determined Eq. \ref{delta} in Table
\ref{deltatab}, are with respect to an inversion center
between two bilayers (point O in Fig. \ref{crystal}(a)).
Thus, for a crystal cleaved between two bilayers,
$N(\Lambda_a)$ can be deduced by combining Eq. \ref{fermionparity} with
\begin{equation}
\pi(\Lambda_a) = - \delta(\Gamma_{a1})\delta(\Gamma_{a2})
\label{pi0}
\end{equation}
as shown in Table \ref{fptab}.  This implies
 the surface Fermi surface encloses $\bar\Gamma$, but not $\bar M$,
 as shown schematically in Fig. \ref{bzfig}(a).
Eq. \ref{fermionparity} says nothing about either the hole pockets
seen in experiment and our calculation or
the double electron pocket at $\bar M$ observed in experiment\cite{hasan} on
Bi$_{1-x}$Sb$_x$ but not our calculation.
In order for the surface to be neutral, however, the Fermi energy must either be at a Dirac
point at $\bar\Gamma$ (so that the Fermi surface has vanishing area) or there must also be
compensating electron/hole pockets elsewhere in the surface Brillouin zone
(but not enclosing $\bar M$).

It is also instructive to first consider a $(111)'$ face cleaved
{\it between} the monolayers in a bilayer, despite the fact that such
a surface would likely be unstable.  Since the origin ${\bf c}_0$ is
not in the cleavage plane, the parity eigenvalues in (\ref{delta}) need to be
modified using \ref{piprime}.
This has the effect of changing the sign of all of the $\pi(\Lambda_a)$, so that
\begin{equation}
\pi'(\Lambda_a) =
+\delta(\Gamma_{a1})\delta(\Gamma_{a2}).
\label{pi1}
\end{equation}
From Table \ref{fptab} we thus conclude that the three $\bar M$ points are enclosed
by the Fermi surface, but not $\bar \Gamma$, as shown in Fig. \ref{bzfig}(b).

For the 110 surface the cleavage plane with one broken bond includes
the origin ${\bf c}_0$.  Thus $\pi(\Lambda_a)$ can be determined
with (\ref{pi0}) along with the projections of the bulk TRIM shown in Table \ref{fptab}.
This leads to the predictions for the surface Fermi surface shown in
Fig. \ref{bzfig}(c).  Experimental data for this face of Bi$_{1-x}$Sb$_x$ is
currently unavailable.  However, it is instructive to compare this
prediction with experiments on pure Bi.
In Ref. \onlinecite{hofmann2}, single hole
pockets are clearly seen at $\bar \Gamma$ and $\bar M$, and at $X_1$
single surface Dirac point is present inside the bulk gap.
The situation at $\bar X_2$ is obscured due to the overlap of the
bulk conduction and valence bands at $L$ and $T$.

For the 100 surface the cleavage plane with one broken bond does not
include ${\bf c}_0$.  Thus, as was the case for the $(111)'$ surface,
the surface fermion parity follows from (\ref{pi1}).  The surface Brillouin
zone shown in Fig. \ref{bzfig}(d) has TRIM $\bar\Gamma$, $\bar M$ and two
equivalent $\bar M'$.   Again, there is presently no data for this surface
of Bi$_{1-x}$Sb$_x$.  The (100) face of pure Bi is discussed in
Ref. \onlinecite{hofmann1}, and appears to be consistent with the prediction
of Table \ref{fptab} that
none of the TRIM are enclosed by a Fermi surface.

\section{Mirror Chern number and the mirror chirality at the L point
of Bismuth}

In this section we will explore the consequences of mirror symmetry
on the band structure of Bi and Bi$_{1-x}$Sb$_x$.
This will address the disagreement between our calculation of the surface
band structure and previous experimental and theoretical results.  As discussed in Section 3,
the tight binding model predicts that the hole pockets enclose Dirac points,
while experiment and
first principles calculations suggest that they do not.  Here we will show that the presence
of this crossing probes a fundamental, but previously unexplored, property of the bulk electronic
structure of Bi.

We will begin by pointing out that the mirror symmetry of the rhombohedral A7 structure leads to
an additional topological structure of the energy bands which we refer to as a mirror Chern
number.  We will then show that the value of this integer in the topological insulator phase
depends on the structure of the nearly degenerate $L_s$ and $L_a$ bands in pure Bi.
We will identify a previously unexplored parameter in the  ${\bf k}\cdot{\bf p}$
theory of Bi, which we refer to as the {\it mirror chirality}.  We will show that the mirror
chirality at the L point in Bi determines the value of the mirror Chern number
in the topological insulator phase of Bi$_{1-x}$Sb$_x$.

We find that the value of the mirror chirality predicted by the Liu Allen tight binding
model\cite{liuallen}
disagrees with the value predicted by a more fundamental calculation by Golin\cite{golin}.
This, combined
with the disagreement with the surface state experiments and first principles calculations
suggests that the Liu Allen tight binding model has a subtle, but topological, error.

\subsection{The Mirror Chern Number}

The Dirac points in the hole pockets in our tight binding
calculation arise because the $\bar\Sigma_1$ and $\bar\Sigma_2$ bands
cross on the line connecting $\bar\Gamma$ and $\bar M$ in Fig. \ref{tbsurface}(b).
This crossing is protected by
the invariance of the Hamiltonian under the mirror operation
${\cal M}(\hat x) = P C_2(\hat x)$ which takes $x$ to $-x$.
$\bar\Sigma_1 (\bar\Sigma_2)$ transform under different representations of
${\cal M}(\hat x)$ with eigenvalues
$+i(-i)$.  This mirror symmetry implies that all the bulk electronic states in the plane $k_x=0$
can be labeled with a mirror eigenvalue $\pm i$.  Within this two dimensional plane in momentum
space, the occupied energy bands for each mirror eigenvalue will be associated with a {\it Chern
invariant} $n_{\pm i}$.  Time reversal symmetry requires that $n_{+i}+n_{-i}=0$, but the
difference defines a non trivial {\it mirror Chern number}
\begin{equation}
n_{\cal M} = (n_{+i}-n_{-i})/2.
\end{equation}
The situation is analogous to the quantum spin Hall state in graphene\cite{km1,km2},
where the conservation
of spin $S_z$ leads to the definition of a spin Chern number\cite{sheng},
whose parity is related to the $\mathbb{Z}_2$ topological invariant.

The mirror Chern number determines how the surface states connect the valence and conduction bands
along the line $q_x=0$ between $\bar\Gamma$ to $\bar M$.  To
see this, consider the ${\cal M} = \pm i$ sectors independently.  The bulk states with $k_x=0$ are
then analogous to a
two dimensional integer quantum Hall state with Hall conductivity $n_{\pm i} e^2/h$.  The sign of
$n_{\pm i}$ determines the direction of propagation of the edge states, which connect the valence and
conduction bands.  Thus, the sign of $n_{\cal M}$ determines whether the $\bar \Sigma_1$ band or the
$\bar\Sigma_2$ band connects the valence and conduction band between $\bar\Gamma$ and $\bar M$
(which we take to be in the $+\hat y$ direction).  For $n_{\cal M} = +1(-1)$
we find that the $\bar\Sigma_1 (\bar\Sigma_2)$ band crosses.

The predictions of the tight binding model are more likely to be robust near $\bar\Gamma$ than
near $\bar M$, because near $\bar\Gamma$ they are not sensitive to the detailed treatment of the
small bulk energy gap at the $L$ point.  This is supported by the fact that the ordering of the
$\bar\Sigma_1$ and $\bar\Sigma_2$ bands near $\bar\Gamma$ predicted by the tight binding model
(in which $\bar\Sigma_2$ emerges first)
agrees with other calculations and experiment.  Given this ordering near $\bar\Gamma$, the mirror
Chern number determines whether or not the $\Sigma_1$ and $\Sigma_2$ bands have to cross.
Referring to Fig. \ref{tbsurface}(c), if the mirror Chern number were to have the opposite sign,
then the
$\bar\Sigma_2$ band would connect to the conduction band rather than the $\bar\Sigma_1$ band, and the
bands would not have to cross.  Pure Bi is very close to the
transition between the $(0;000)$ and $(1;111)$ phases.  Therefore, it
is likely that the presence of the crossing between $\bar\Sigma_1$
and $\bar\Sigma_2$ will be unaffected by the transition.  Therefore,
the sign of the $n_{\cal M}$ in the topological insulator phase of
Bi$_{1-x}$Sb$_x$ should be correlated with the alternatives shown in
Figs. \ref{tbsurfacebi}, with $n_{\cal M} = + (-) 1$ corresponding to
Fig. \ref{tbsurfacebi}b (\ref{tbsurfacebi}c).

Since the valence band of pure Bi is in the trivial $(0;000)$ topological class, pure Bi
does not have surface states which connect the valence and conduction bands.  Thus the mirror
Chern number for the $k_x=0$ plane of the valence band of pure Bi is $n_{\cal M}=0$.   The
transition to the strong topological insulator in Bi$_{1-x}$Sb$_x$ occurs for small $x$ because
the $L_s$ and $L_a$ bands in pure Bi are nearly degenerate.  At the transition to the
topological insulator the two bands cross and form a three dimensional Dirac point at $L$.
At this transition both the $\mathbb{Z}_2$ topological invariants $(\nu_0;\nu_1\nu_2\nu_3)$ and the
mirror Chern number $n_{\cal M}$ change.  The {\it change} $\Delta n_{\cal M}$ across this transition is an
intrinsic property of this Dirac point.  Thus the value of $n_{\cal M}$ in the topological insulator
phase can be determined by studying the properties of this Dirac point.  Since pure Bi is
very close to this transition, this information can be extracted from the structure of the
${\bf k}\cdot{\bf p}$ Hamiltonian for pure Bi in the vicinity of the $L$ point.

In the next section we will analyze the ${\bf k}\cdot{\bf p}$ theory and
 show that the value of $\Delta n_{\cal M}$ predicted by the Liu Allen tight binding
model {\it disagrees} with the value predicted by an earlier pseudopotential calculation by Golin.
This provides evidence that the crossing
of the $\bar\Sigma_1$ and $\bar\Sigma_2$ bands is an artifact of the incorrect sign of $n_{\cal M}$
 predicted by the tight binding model.

\subsection{${\bf k}\cdot{\bf p}$ theory and the mirror chirality}

The ${\bf k} \cdot {\bf p}$ analysis of Bi near the L point has a long history.  Originally
developed by Cohen and Blount\cite{cohen} in 1960, the theory was given an particularly elegant
formulation by Wolff\cite{wolff}, who emphasized the similarity with the relativistic Dirac equation.
This
theory, and its refinements\cite{smith,baraff,buot} played an important role in the
early development of band theory, and
formed the framework for interpreting a large body of magnetic, transport and optical data.
In this section we point out a previously unexplored sign which characterizes this theory : the
mirror chirality.  We show that it is this sign which determines the sign of $n_{\cal M}$ in the
topological insulator phase.

The four relevant states at the $L$ point are denoted $[L_s,L_a] = [(L_6,L_5),(L_7,L_8)]$\cite{falikov}.
The two states comprising $L_s$ and $L_a$ are degenerate due to time reversal symmetry.
These states are distinguished by their symmetry under parity $P$ (with eigenvalues
$[(1,1),(-1,-1)]$), under the twofold rotation $C_2(\hat x)$ (with eigenvalues $[(-i,i),(i,-i)]$)
and under the mirror ${\cal M}(\hat x) = P C_2(\hat x)$
(with eigenvalues $[(-i,i),(-i,i)]$).  We have chosen the
unconventional order of the states to simplify the mirror operator, which makes
the connection with the mirror Chern number in section VC the most transparent.   In this basis the
inversion, rotation and mirror operators have the direct product form,
\begin{eqnarray}
P &=& \tau_z\otimes\openone, \nonumber\\
 C_2(\hat x) &=& -i \tau_z\otimes\mu_z, \\
{\cal M}(\hat x) &=& -i \openone\otimes\mu_z, \nonumber
\end{eqnarray}
while the time reversal operator can be chosen as
\begin{equation}
\Theta = i \openone\otimes\mu_y K,
\end{equation}
where $K$ is complex conjugation.  $\vec \mu$ and $\vec \tau$ are Pauli matrices operating
within and between the $L_s$ and $L_a$ blocks, and $\openone$ is the identity matrix.
In the following we will simplify the notation by omitting the $\otimes$ and the $\openone$.

To first order in ${\bf k}$ the ${\bf k}\cdot {\bf p}$ Hamiltonian has the form
\begin{equation}
H({\bf k}) = m \tau_z + k_x \Pi_x + k_y \Pi_y + k_z \Pi_z,
\label{hkp}
\end{equation}
Where $E_G = 2m$ is the energy gap (positive for Bi) and $\Pi_a$ are
$4\times 4$ matrices.  Invariance of $H({\bf k})$ under $P$ and $\Theta$ requires $\{\Pi_a,P\}=\{\Pi_a,\Theta\}=0$, and
invariance under ${\cal M}(\hat x)$ requires $\{\Pi_x,{\cal M}(\hat x)\}=[\Pi_{y,z},{\cal M}] = 0$.  The allowed
terms are thus
\begin{eqnarray}
&\Pi_x = t_1 \tau_x  \mu_x + t_2 \tau_x  \mu_y \nonumber\\
&\Pi_y = u_{11} \tau_x  \mu_z + u_{12} \tau_y   \\
&\Pi_z = u_{21} \tau_x  \mu_z + u_{22} \tau_y \nonumber
\label{pixyz}
\end{eqnarray}
where $t_i$ and $u_{ij}$ are real numbers.
Eqs. \ref{hkp} and \ref{pixyz} are equivalent to the ${\bf k}\cdot{\bf p}$ theory
introduced by Cohen and Blount\cite{cohen},
who expressed the Hamiltonian in terms of  complex vectors ${\bf t}$ and ${\bf u}$.
These are related to our parameters via
${\bf t} = (t_1 + i t_2)\hat x$ and
${\bf u} = (-u_{11} + i u_{12})\hat y + (-u_{21} + i u_{22})\hat z$.  In the following it will be
useful to express these in terms of three complex numbers
$t = \hat x\cdot{\bf t}$ and $u^{\pm} = (\hat y \pm i \hat z) \cdot {\bf u}$.

Eq. \ref{hkp} has a simpler form when expressed in terms of the principle axes in both momentum space and
spin space.  We thus perform a rotation $(k_y+ i k_z) = e^{i\alpha} (k'_y + i k'_z)$ along with a
unitary transformation $|\psi\rangle = \exp[ i \mu_z (\beta + \gamma \tau_z)] |\psi'\rangle$.
These transformations have the effect of changing the phases $t\rightarrow t e^{i\beta}$ and
$u^{\pm} \rightarrow u^{\pm} e^{-i(\gamma \pm \alpha)}$.
For appropriately chosen $\alpha$, $\beta$ and $\gamma$, $t$ and $\mp u^{\pm}$ can be
made real and positive.  The Hamiltonian then takes the diagonal form
\begin{equation}
H = m \tau_z + v_1 k_x \tau_x \mu_x + \eta v_2 k'_y \tau_x \mu_z + v_3 k'_z \tau_y.
\label{hdiagonal}
\end{equation}
where
\begin{eqnarray}
v_1 &=& |t|  \nonumber \\
\eta v_2 &=& (|u^+| - |u^-|)/2 \\
v_3 &=& (|u^+| + |u^-|)/2. \nonumber
\end{eqnarray}
Here we have defined $v_2$ to be positive and introduced
a previously unexplored quantity $\eta = \pm 1$, which is simply given by
$\eta=  {\rm sgn}(\det[u_{ij}])$.
$\eta$ is a {\it mirror chirality}, which distinguishes two
topologically distinct classes of Dirac Hamiltonians.

For a system with full rotational symmetry, $\eta$ must be equal to $+1$.  This can be seen by
noting that the twofold rotation operator specifies the generator of
continuous rotations about $\hat x$ via $C_2(\hat x) = \exp[-i\pi S_x]$.
Since $C_2(\hat x) = -i \mu_z\tau_z$, this implies $S_x = \mu_z\tau_z/2$.
When $\eta=-1$, Eq. \ref{hdiagonal} is {\it not} invariant under continuous rotations
generates by $S_x$ even when $v_2 = v_3$, since the spin and orbital degrees of freedom are
rotated in opposite directions.  The twofold rotational symmetry, however, remains intact.
$\eta=+1$ corresponds to the behavior of a free electron and
should be considered normal behavior.  $\eta=-1$ is
anomalous.

The sign of $\eta$ is not ordinarily discussed in the ${\bf k}\cdot{\bf p}$ theory of Bi
because it has no
effect on the electronic dispersion $E({\bf k})$, which depends only on $|v_a|$.
$\eta$ does, however, have a subtle effect in the presence of a magnetic field.  A magnetic field
in the $\hat x$ direction leads to a splitting of states according to their spin angular momentum
$S_x$, which can be defined as above in terms of the twofold rotation operator $C_2(\hat x)$.
This defines a magnetic moment, which symmetry restricts to be either parallel or antiparallel to $\hat x$.
The form of this magnetic moment is discussed in Refs. \onlinecite{wolff,smith}, and it is straightforward to show that
$\vec\mu \propto \eta S_x \hat x$.  This means the $\eta$ determines the {\it sign} of the $g$
factor, which describes the relation between the magnetic moment and angular momentum.  For
$\eta=+1$ the sign is the same as that for a free electron, while $\eta=-1$ the sign is opposite.

Unfortunately, this sign is difficult to probe experimentally.  In addition to complications which arise
due to the presence of three equivalent $L$ points, measurement of the sign requires measurement of
the spin angular momentum in addition to the change in energy with magnetic field.  The selection
rules discussed in Ref. \onlinecite{wolff} are unaffected by the sign.  We are not aware of any experiments on
Bi which directly probe this sign.

\subsection{Relation between mirror chirality, mirror Chern number and surface states}

We will now argue that the sign of $\eta$ determines the sign of the mirror Chern number in the
topological insulator phase of Bi$_{1-x}$Sb$_x$.  This leads to an experimentally testable
prediction regarding the crossing of the surface states.
Thus, probing the surface states of the topological insulator may
well be the best experimental method for determining this fundamental parameter of the
${\bf k}\cdot {\bf p}$ theory of Bi.

The connection between $\eta$ and the mirror Chern number can be established by
considering the mirror plane $k_x=0$.   $H$ then decouples into two independent
two band Hamiltonians for ${\cal M}(\hat x) = - i\mu_z = \pm i$ with the form
\begin{equation}
h  = m \tau_z +  s v_2  k'_y \tau_x  + v_3 k'_z \tau_y .
\label{h2band}
\end{equation}
where $s = \eta\mu_z$.  $m=0$ describes a transition where the Chern number $n_{-i\mu_z}$ changes.
When $m$ changes sign from negative to positive, $\Delta n_{-i\mu_z} = \eta \mu_z$.  Thus, the
change in the mirror Chern number,
\begin{equation}
\Delta n_{\cal M} = n_{\cal M}(m>0) - n_{\cal M}(m<0) = -\eta,
\end{equation}
depends on the mirror chirality $\eta$.  Since $n_{\cal M}=0$ for Bi (with $m>0$),
we conclude that the topological insulator, with $m<0$ has
\begin{equation}
n_{\cal M} =  \eta.
\end{equation}

$n_{\cal M}$ determines
the direction of propagation of the $\bar\Sigma_1$ and $\bar\Sigma_2$ surface states along the
mirror line $q_x=0$.
The direction of propagation of the surface states on the top surface
which connect the valence and conduction
bands can determined by solving (\ref{h2band}) with a $z$
dependent mass $m(z) = m {\rm sign}(z)$ with $m>0$.  The bound state
at the surface has wavefunction
proportional to $\exp(- |m z|/v_3 )$.  The
dispersion for the surface states on the top surface along $q_x=0$ is
\begin{equation}
E(q_y) = - \eta \mu_z v q_y
\end{equation}
with $v>0$.  This means that the $\Sigma_1$ band, which has $\mu_z = -1$,
propagates in the $+\eta \hat y$ direction, while the $\Sigma_2$ band, with $\mu_z = +1$
propagates in the
$-\eta\hat y$ direction.  Therefore, the surface state connecting the valence band to the
conduction band which has the {\it positive} velocity in the $\hat y$ direction will be
$\Sigma_1$ for $\eta = +1$ and $\Sigma_2$ for $\eta = -1$.

\subsection{Comparison of tight binding and pseudopotential models with experiment}

In this subsection we show that the value of $\eta$ predicted by the Liu Allen tight binding
model\cite{liuallen}
disagrees with that predicted by an early calculation by Golin\cite{golin}.  Specifically,
we find that the
Liu Allen model predicts the conventional value, $\eta=1$, while the Golin model predicts the
anomalous value $\eta = -1$.   We will then argue that the value
of $\eta$ can be extracted from the structure of the surface state spectrum.  The presently
available spin polarized ARPES data on the Bi 111 surface\cite{hirahara2} provides
indirect evidence that the mirror chirality has the anomalous value $\eta = -1$.

The ${\bf k}\cdot{\bf p}$ parameters can be determined by evaluating the matrix elements
\begin{equation}
\Pi_a^{ij} = \langle L_i |\hat v_a| L_j \rangle|_{{\bf k}=L}.
\end{equation}
where ${\bf\hat v} = \nabla_{\bf k}{\cal H}({\bf k})|_{{\bf k}=L}$ is determined by the Bloch Hamiltonian
${\cal H}({\bf k})$.  From this it follows that
\begin{eqnarray}
t &=& \Pi_x^{57} \\
u^\pm &=& -\Pi_y^{67} \mp i \Pi_z^{67}.
\end{eqnarray}
These matrix elements are listed in table II of Golin's paper\cite{golin} (the relevant band is $j=j'=3$).
They may also be extracted from the Liu Allen tight binding model.
In Table \ref{kptable} we compare the values of $v_1$, $v_2$, $v_3$ and $\eta$
computed from these matrix elements.  The signs of $\eta$ predicted by the two theories disagree.
Since the parameters of the Liu
Allen model were simply fit to reproduce the {\it energies} of the bands, there is no reason to expect
that it gets $\eta$ right.  In contrast, Golin's calculation, which is based on a pseudopotential
approach, starts from more fundamental premise.

\begin{table}[htbp] \begin{center}
 \begin{tabular}{|c|c|c|c|c|}
 \hline
 { } & $v_1$ (eV\AA) & $v_2$ & $v_3$ & $\eta$ \\
\hline
Golin Pseudopotential& $4.16$ & $1.37$ & $7.01$ & $-1$ \\
\hline
Liu Allen Tight Binding & $5.89$ & $0.92$ & $9.67$ & $+1$ \\
\hline
\end{tabular} \label{tab1} \end{center}
\caption{Parameters of the ${\bf k}\cdot{\bf p}$ theory, Eq. \ref{hdiagonal}, extracted
from the pseudopotential model\cite{golin} and the tight binding
model\cite{liuallen}}.
\label{kptable}
\end{table}

In the previous section we showed that
provided there is only a single transition between pure Bi and the topological insulator
phase of Bi$_{1-x}$Sb$_x$, the mirror chirality deduced from the pure Bi band structure
determines the mirror Chern number in the topological insulator.  This, in turn, determines
the direction of propagation of the $\bar\Sigma_1$ and $\bar\Sigma_2$ states along the line
$q_x=0$.  The surface state structure predicted by the tight binding model was shown in Fig.
\ref{tbsurface}(c).  The crossing of the $\Sigma_1$ band is consistent with $\eta = +1$.
This crossing guarantees that there is a Dirac point enclosed by the hole pocket.
This can be probed either by inverse photoemission or by spin polarized photoemission.
In the latter case, the presence of the Dirac point would lead to a change in the sign of the
spin on either side of the hole pocket.  It will be interesting to experimentally determine
this property for Bi$_{1-x}$Sb$_x$ using spin polarized ARPES.

Currently available spin polarized photoemission data on the 111 surface of pure Bi\cite{hirahara2}
 provide an indirect probe of $\eta$.  Hole pockets are observed
along the line from $\bar\Gamma$ to $\bar M$ in both Bi$_{1-x}$Sb$_x$ and pure Bi.
Provided we make the plausible assumption that no additional level crossings occur near the
transition to the topological insulator, then the presence or absence of Dirac points in
the hole pockets should be the same on both sides of the transition.  In Ref. \onlinecite{hirahara2},
the spin in
either side of the hole pocket was found to point in the same direction, which indicates that in
pure Bi, the hole pockets do not enclose a Dirac point.  This conclusion was supported by
first principles surface state calculations, which also find no crossing\cite{hirahara2}.
This suggests that in the
alloy, it should be the $\Sigma_2$ band which connects the conduction and valence bands,
which is consistent with $\eta = -1$.

It thus appears likely that the mirror chirality in Bi has the anomalous sign, $\eta = -1$.
This conclusion contradicts the prediction of the tight binding model, but it is supported by
(1) the pseudopotential band structure of pure Bi and (2) the observed and calculated surface
state structure of pure Bi.  Spin polarized ARPES experiments on the topological
insulator Bi$_{1-x}$Sb$_x$ could more directly determine this sign by probing the mirror Chern number
$n_{\cal M}$.

\section{Conclusion}

In this paper we have analyzed the surface state structure of the topological insulator
Bi$_{1-x}$Sb$_x$.  Using a simple tight binding model based on Liu and Allen's tight binding
parameterization we confirmed that the surface states have the signature of the strong topological
insulator by showing that the surface Fermi surface encloses an odd number of Dirac points.  The
tight binding model also predicts that the surface is semi metallic, with an electron pocket
centered on $\bar\Gamma$ along with 6 hole pockets.

Using general arguments based on inversion symmetry, we showed that the location of electron and
hole pockets in the surface Brillouin zone is constrained by a quantity which we defined as the
surface fermion parity.  This quantity is determined by
the parity invariants of the bulk band structure, and for a given surface
it determines which surface TRIM are enclosed by an
odd number of Fermi surface
lines.  This argument establishes a simple and direct connection between the bulk electronic
structure and the surface electronic structure for crystals with inversion symmetry.
Using this general principle, we predicted the structure of the surface states for several
different faces of Bi$_{1-x}$Sb$_x$.  For the 111 face, these predictions agree both with our
surface state calculations and with experiment.  It will be interesting to test these predictions
experimentally on other faces of Bi$_{1-x}$Sb$_x$.

Finally, we showed that the mirror symmetry present in the rhombohedral A7 lattice leads to
additional topological structure in the bulk energy bands.  We defined an integer
mirror Chern number $n_{\cal M}$, whose value is nonzero in the topological insulator phase.  The sign of
$n_{\cal M}$ determines the direction of propagation of each of the surface states along the mirror
plane, and thus determines which surface states connect the conduction and valence bands.  We find
that the crossing of the $\Sigma_1$ band predicted by the tight binding model, which
leads to a Dirac point in the hole pockets, disagrees with
the natural extrapolation of experiments and first principles calculations on pure Bi, which
find no Dirac point in the hole pockets.

We traced this discrepancy to a previously unexplored property of the ${\bf k}\cdot{\bf p}$ band
structure of pure Bi, which we defined as the mirror chirality, $\eta$.  We showed that
$\eta$ in pure Bi determines $n_{\cal M}$ in the topological insulator.  Moreover, we showed that
the Liu Allen model predicts the conventional value
$\eta = +1$, while an earlier pseudopotential calculation by Golin
predicts the anomalous value $\eta = -1$.  The latter value is consistent with the available
experimental data on Bi, though the connection is rather indirect.  A more direct test would
be to directly measure the mirror chirality $n_{\cal M}$ in the topological insulator by probing the
surface states with spin polarized ARPES.

It would be interesting to check that the value of $\eta$ predicted by more accurate first principles
calculations of Bi agrees with the pseudopotential prediction.
Since the tight binding model was designed only to get the energies of the bands right, there is
no reason to expect that it would get $\eta$ right.  It should be possible to come up with a new
parameterization of the Liu Allen model which would have $\eta = -1$.  We expect that the surface
states computed within this model would have band crossings which agree with experiment and first
principles calculations, though of course a quantitative description of the surface states
requires an accurate description of the surface potential.

An important lesson to be learned from this paper is that in addition to time reversal symmetry,
spatial symmetries can play an important role in topologically constraining bulk and surface
band structures.  Our analysis of these symmetries has not been exhaustive.  A complete theory of
{\it topological band theory}, which accounts for the full point group symmetry of a crystal
is called for.

\acknowledgments

We thank Gene Mele for helpful discussions and Zahid Hasan and
David Hsieh for sharing their experimental results prior to publication.
This work was supported by NSF grant DMR-0605066, and by ACS PRF grant 44776-AC10.

\begin{appendix}

\section{Surface Fermion Parity from Bulk Parity Invariants}

In this appendix we show that for an inversion and time reversal invariant
crystal the surface fermion number $N({\bf q}=\Lambda_a)$
discussed in section IV is an integer, whose parity is determined by the
product of bulk parity invariants $\delta(\Gamma_{a1,2})$, which are
products of parity eigenvalues given
in Eqs. \ref{delta} and \ref{pi}.   The simple proof
outlined here provides a direct connection between the topological
structure of the surface states and the parity eigenvalues
characterizing the bulk crystal.

\begin{figure}
\centerline{
\epsfig{figure=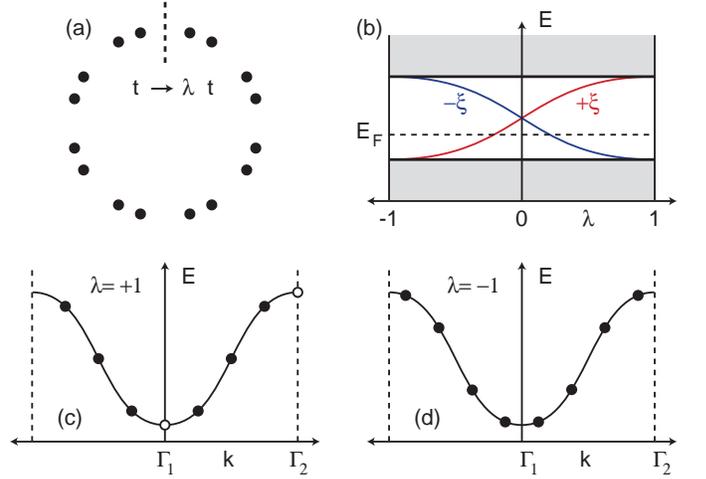,width=3.5in} }
 \caption{(a) A one dimensional inversion symmetric insulator cut at $z=0$ by replacing
 hopping amplitudes $t$
across $z=0$ by $\lambda t$. The fully cleaved crystal corresponds to $\lambda=0$.
(b)  Energy spectrum as a function of $\lambda$ between $-1$ and $1$.
The conduction and valence bands
exchange a Kramers pair of states with opposite parity.
(c,d) The bulk energy levels at $\lambda=\pm 1$. For $\lambda=-1$ (d) every
state at $k$ has a partner at $-k$ with the same energy and opposite parity.
For $\lambda=+1$ (c) the states at $k=\Gamma_1$ and $k=\Gamma_2$ are not paired. }
 \label{1dfig}
\end{figure}

The Bloch Hamiltonian $H(\Lambda_a,k_z)$
describes a parity and time reversal invariant one dimensional system.
In the following we will suppress the dependence on $\Lambda_a$ and
consider a purely one dimensional system.  To determine the end charge
$N$ we introduce the ``cutting procedure"
depicted in Fig. \ref{1dfig}(a).  We begin with a large but finite system with periodic
boundary conditions.  We then replace the hopping amplitudes $t_i$ for all
bonds that cross the cleavage plane $z=0$ by $\lambda t_i$, where
$\lambda$ is real.  Provided $z=0$ corresponds to an
inversion plane, the one dimensional Hamiltonian retains inversion
and time reversal symmetry for all $\lambda$.  The fully cleaved
crystal corresponds to $\lambda = 0$.

For $\lambda=1$ the system is translationally invariant, so the
excess charge near $z=0$ is $Q(\lambda=1)=0$. Since
the insulator can have no bulk
currents, the only way $Q(\lambda)$ can change is if a state
localized near $z=0$ crosses the Fermi energy.  Thus $Q(0)$
will be the difference between the number of states that cross $E_F$
from above and from below for $\lambda \in [0,1]$.  Kramers' theorem
requires that every state is at least twofold degenerate, so the
number of states crossing $E_F$ will be an {\it even} integer.  Since
the charge will be divided evenly between the two sides, $N =
Q(0)/2$ is an integer, which may be written
\begin{equation}
N = \Delta N_+ - \Delta N_-,
\end{equation}
where
$\Delta N_\pm$ is the number of {\it Kramers' pairs} that cross $E_F$
from above or below.

We now relate the parity of $N$
to the bulk parity eigenvalues.  To this end it is useful to consider
the evolution of the spectrum for $\lambda\in[-1,1]$ and to define
\begin{equation}
P(\lambda) = \prod_{E_{2\alpha}(\lambda) < E_F} \xi_{2\alpha}
\label{pprod}
\end{equation}
 as the product of the parities of all of the occupied
states, where each Kramers pair $(\psi_{2\alpha},\psi_{2\alpha-1})$
is included only once. This quantity is well defined because
$\xi_{2\alpha}=\xi_{2\alpha-1}$.  Our proof consists of two steps.
We will first show that
\begin{equation}
P(1)P(-1) = (-1)^N.
\label{pp1}
\end{equation}
We will then show that
\begin{equation}
P(1) P(-1)  =
\prod_{m=1}^{n_b} \left[- \xi_{2m}(\Gamma_1) \xi_{2m}(\Gamma_2)
\right] \equiv \pi.
\label{pp2}
\end{equation}
Here $\xi_{2m}(\Gamma_i)$ are the parity
of the Bloch states in the $m$th Kramers degenerate
band at the TRIM $k_z = \Gamma_i$, and
again each Kramers pair is included only once.  $n_b$ is the number
of occupied Kramers degenerate bands.  Taken together, (\ref{pp1}) and (\ref{pp2})
establish the relationship summarized by Eqs. \ref{delta} and \ref{pi}  between the
bulk parity eigenvalues and the surface fermion parity.

Eq. \ref{pp1} follows from the symmetry of the end state spectrum about $\lambda =0$.
The Hamiltonian $H(-\lambda)$ differs $H(\lambda)$
only by a phase twist of $\pi$ across $z=0$.
This twist can be spread over the entire circumference $L$ by performing
the gauge transformation
\begin{equation}
|\psi(-\lambda)\rangle = e^{i \pi z/L}|\tilde\psi(-\lambda)\rangle
\label{gauge}
\end{equation}
for $0<z<L$.  When $L\rightarrow\infty$ the Hamiltonian for
$|\tilde\psi(-\lambda)\rangle$ near
$z=0$ becomes identical to $H(\lambda)$.  Thus every bound state
$|\psi_l(\lambda)\rangle$ satisfies $E_l(-\lambda) = E_l(\lambda)$.
Since (A5) changes the parity,
$|\psi_l(\lambda)\rangle$ and  $|\psi_l(-\lambda)\rangle$ have
opposite parity.

It follows that every Kramers pair that crosses the $E_F$
at $\lambda_0 \in [1,0]$ has a partner
with opposite parity that crosses $E_F$ in the opposite direction
at $-\lambda_0$ as shown in Fig \ref{1dfig}(b).  Thus between $\lambda=1$ and
$\lambda=-1$
the conduction and valence band exchange two Kramers pairs with opposite
parity, leading to a change in the relative
sign between $P(1)$ and $P(-1)$.   We conclude that
$P(1)P(-1) = (-1)^{\Delta N_+ + \Delta N_-}$, which leads directly to
(\ref{pp1}).

Eq. \ref{pp2} follows from a consideration of the parities of the Bloch wavefunctions.  Consider first
the simplest case where there is a single Kramers degenerate
occupied band, as shown in Fig. \ref{1dfig}(c,d). At $\lambda=1$
the single particle states are labeled by momentum $k_z=2m\pi/L$
with $m=-M/2+1, ..., M/2$, where
$M$ is the number of unit cells. At the two TRIM $\Gamma_1 = 0$, $\Gamma_2= M\pi/L$ the parity
eigenvalues are $\xi(\Gamma_{1,2})$.
Every other $k_z$ has a partner $-k_z$,
and even and odd parity combinations of the two can be formed.  The $M/2-1$
$(k_z,-k_z)$ pairs thus each contribute $-1$ to the product in (\ref{pprod}). Therefore,
\begin{equation}
P(1)= (-1)^{M/2-1}\xi(\Gamma_1)\xi(\Gamma_2).
\label{p1}
\end{equation}
For $\lambda=-1$ the gauge transformation (\ref{gauge}) leads to
a periodic Hamiltonian identical to $H(1)$, but with momenta shifted
by $\pi/L$, as shown in Fig. \ref{1dfig}(d).  Thus all the momenta are paired,
so that
\begin{equation}
P(-1)=(-1)^{M/2}.
\label{p2}
\end{equation}
Combining (\ref{p1}) and (\ref{p2}) leads directly to (\ref{pp2}), which is straightforwardly
generalized to the case of $n_b$ Kramers degenerate bands.

\end{appendix}

\end{document}